\documentclass[journal=ancac3,manuscript=article]{achemso}

\usepackage{chemformula} 
\usepackage[utf8]{inputenc} 
\usepackage[T1]{fontenc} 
\usepackage{graphicx} 
\usepackage{wrapfig}
\usepackage{lineno}

\author{Adrian Hemmi}
\affiliation{Physik-Institut, Universität Zürich, 8057 Zürich, Switzerland}
\altaffiliation{Contributed equally to this work}

\author{Huanyao Cun}
\affiliation{Physik-Institut, Universität Zürich, 8057 Zürich, Switzerland}
\altaffiliation{Contributed equally to this work}

\author{Steven Brems}
\affiliation{IMEC vzw Kapeldreef 75, 3001 Leuven, Belgium}
\author{Cedric~Huyghebaert}
\affiliation{IMEC vzw Kapeldreef 75, 3001 Leuven, Belgium}

\author{Thomas Greber}
\email{greber@physik.uzh.ch}
\phone{+41 44 635 57 44}
\affiliation{Physik-Institut, Universität Zürich, 8057 Zürich, Switzerland}

\title{Wafer-scale, epitaxial growth of single layer hexagonal boron nitride on Pt(111)}

\keywords{Hexagonal boron nitride, CVD growth on catalysts, in-situ photoelectron yield measurements, growth kinetics, wafer scale}

\begin{document}









\begin{abstract}
Single layer hexagonal boron nitride is produced on 2 inch Pt(111)/sapphire wafers. The growth with borazine vapour deposition at process temperatures between 1000 and 1300~K is in-situ investigated by photoelectron yield measurements.
The growth kinetics is slower at higher temperatures and follows a $\tanh^2$ law which better fits for higher temperatures.
The crystal-quality of h-BN/Pt(111) is inferred from scanning low energy electron diffraction (x-y LEED).
The data indicate a strong dependence of the epitaxy on the growth temperature.
The dominant structure is an aligned coincidence lattice with 10 $h$-BN on 9 Pt(1$\times$1)  unit cells and follows the substrate twinning at the millimeter scale. 
\end{abstract}

\section{Introduction}

Two-dimensional (2D) materials are expected to become important for the application of concepts for platforms and building blocks in electronic  devices beyond the silicon technology \cite{Novoselov2004, Banszerus2015}, in nanophotonics \cite{Tran2015}, in membrane technology \cite{Surwade2015}, for mechanical detectors \cite{Cartamil-Bueno2017} or as interface systems in electrochemical environments \cite{Mertens2016}. 
All these promises rely on the production of the materials at large scale and in sufficient quality. Quality as a degree of excellence may have several aspects, like crystallinity, homogeneity, stoichiometry, defect density etc., after growth, transfer or processing.  
In the present contribution to this focus on hexagonal boron nitride ($h$-BN) we report advances and challenges of epitaxial growth of single layer $h$-BN with chemical vapour deposition (CVD) in ultra-high vacuum (UHV) environment and pick platinum as a prototype growth substrate. The crystal-quality after growth is quantified from low energy electron diffraction patterns.

There are excellent reviews on CVD grown single layer $h$-BN, covering applications in surface science \cite{Auwarter2019} and beyond \cite{Zhang2017}. 
First reports on the adsorption of borazine on Pt(111) and Ru(0001) surfaces date back to 1990: Paffett {\it{et al.}} reported 
ordered close-packed $h$-BN overlayers with saturation BN coverages corresponding to one monolayer of $h$-BN \cite{Paffett1990}.
Work with other precursor molecules like trichloroborazine \cite{Auwarter2004} revealed, compared to the 10 on 9 coincidence lattice \cite{Paffett1990} a low energy electron diffraction (LEED) pattern rotated by 30 degrees \cite{Muller2005}.
The $h$-BN/Pt(111) system was further investigated with low energy electron microscopy (LEEM) that gave direct insight into the growth kinetics which is based on a nucleation and growth mechanism \cite{Cavar2008}.

For many applications $h$-BN has to be transferred from the growth substrate, where exfoliation from Pt was demonstrated in 2013 \cite{Gao2013}. 
The choice of the optimal substrate for growth and exfoliation depends on the bond strength of BN to the substrate, where strong bonding favours the single crystallinity of the $h$-BN, though strong bonding might be a disadvantage if it comes to exfoliation.
The dependence of the $h$-BN bond strength to a given substrate \cite{Laskowski2008} and the registry of the BN \cite{Grad2003} was predicted by density functional theory (DFT). E.g. comparison between Pt and Ru indicates a weaker binding of BN to platinum \cite{Laskowski2008,Laskowski2010}, though it appears more difficult to grow single orientation $h$-BN on Pt \cite{Preo2007} as compared to Ru \cite{Goriachko2007}.

Upscaling is a further requirement for material production and was first addressed with the use of single crystalline films as growth substrates \cite{Gsell2009}. This opened the perspective on high quality single orientation layers of $h$-BN on the four inch wafer scale \cite{Hemmi2014}. Here we apply Pt(111) on 2 inch wafers as substrates and optimize the growth conditions.
A high-temperature growth route facilitates diffusion and stitching of coalescing islands, while using a low pressure of borazine minimizes nucleation seed densities and is necessary to promote systems with minimal residual defect numbers \cite{Lu2013}. 
The $h$-BN quality is judged from in-situ monitoring of the growth kinetics at different temperatures with photoelectron yield measurements {\it{and}} the crystal-quality of LEED patterns that are recorded after growth.

\section{Methods}
\subsection {Preparation of the Pt(111)/sapphire wafers}
The platinum thin film preparation follows a scheme reported previously \cite{Verguts2017}. Double polished, epi-ready, c-plane, two-inch sapphire wafers were cleaned in H$_3$PO$_4$/H$_2$SO$_4$ (1:3) at 200$^\circ$C for 20 min and rinsed in ultra pure water for 3 min in an overflow rinse bath. The sapphire substrates were dried using the Marangoni drying effect. Next, a 500~nm thick Pt layer was deposited on sapphire using e-beam evaporation at 550$^\circ$C with an evaporation rate of 0.02-0.03~nm/s. The base pressure of the deposition tool was about $9\times 10^{-7}$ mbar.

\subsection {Growth and characterization of h-BN/Pt(111)}

The single layer $h$-BN is grown in an ultra-high vacuum, chemical vapour deposition (CVD) cold wall chamber for wafers up to 4-inch and was described previously \cite{Hemmi2014}. Prior to all {\mbox{$h$-BN}} preparations the Pt/sapphire substrate was cleaned by argon sputtering, O$_2$ exposure and annealing cycles to 1123~K until sharp Pt(111) ($1\times1$) LEED patterns were observed. Subsequently, $h$-BN was prepared at different substrate temperatures above 1000~K with borazine (HBNH)$_3$ as precursor gas at a partial pressure reading of 10$^{-7}$ mbar. At these conditions single layer growth without hydrogen incorporation is expected. X-ray photoelectron spectroscopy (XPS) on 10$\times$10 mm$^2$ samples cut from the 2-inch wafers confirm this  and for the presented growth parameters stoichiometric single layer boron nitride was found. The sample temperature was measured with a two-color pyrometer (Maurer, QKTRD 1075). The growth kinetics was traced with photoelectron yield measurements with a xenon flash lamp \cite{Hemmi20142}, where photoelectron pulses on an area of 1 cm$^2$ were detected at a rate of about 0.6~Hz with a Keithley 2635A source-measure unit. After growth the structures were investigated in the same apparatus with an OCI BDL600-MCP LEED optics where the diffraction patterns were recorded with a (PCO AG) CCD camera. The cycle time for moving the x-y stage and taking an image was 1.2+0.3 s for positioning and CCD exposure.

\section{Results and Discussion}
\subsection{$h$-BN/Pt(111) growth kinetics}

\begin{figure}
	\includegraphics[width = 0.6\textwidth]{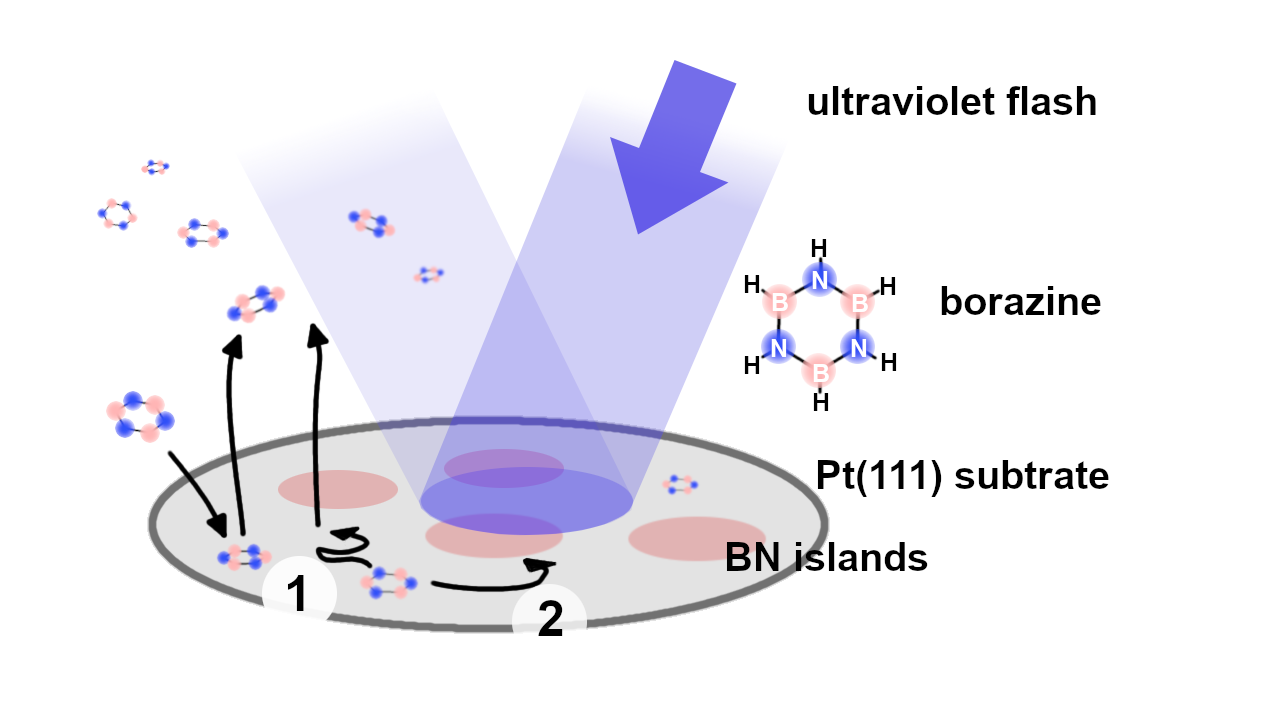}
  \caption{{\bf{ $h$-BN growth process and coverage measurement with photoemission}} An ultraviolet light flash excites photoelectrons, where $h$-BN islands grow during exposure of borazine from the gas phase. A growth kinetics model must include a sticking coefficient $s$, where subsequently BN precursor material may desorb again after some diffusion time (1) or where it contributes to $h$-BN growth (2). The $h$-BN coverage is inferred from the detected photo-charge per light pulse.}
  \label{F0}
\end{figure}

The CVD growth of $h$-BN on transition metals proceeds via nucleation and growth, as it was observed with scanning tunneling microscopy (STM) for $h$-BN/Ni(111) \cite{Auwarter2003} or Rh(111) \cite{Frenken2010} and with LEEM for $h$-BN/Pt(111) \cite{Cavar2008} or Ru(0001) \cite{Sutter2011}.  
The $h$-BN layer grows from homogeneously distributed nucleation centers. Later, the corresponding $h$-BN islands coalesce under further precursor gas exposure to a closed layer. 
Here we apply photoelectron yield measurements \cite{Hemmi20142} for the direct recording of the growth kinetics. Compared to STM or LEEM this method is low cost and more robust in view of temperature and pressure compatibility.
The in-situ surface probing and an overview of the growth process is sketched in Figure \ref{F0}. An ultraviolet light flash excites photoelectrons, which are collected with a biased antenna, where the detected electron yield $Y$ is directly related to the {\mbox{$h$-BN coverage $\Theta$}}. 

\begin{figure}
	\includegraphics[width = 0.6\textwidth]{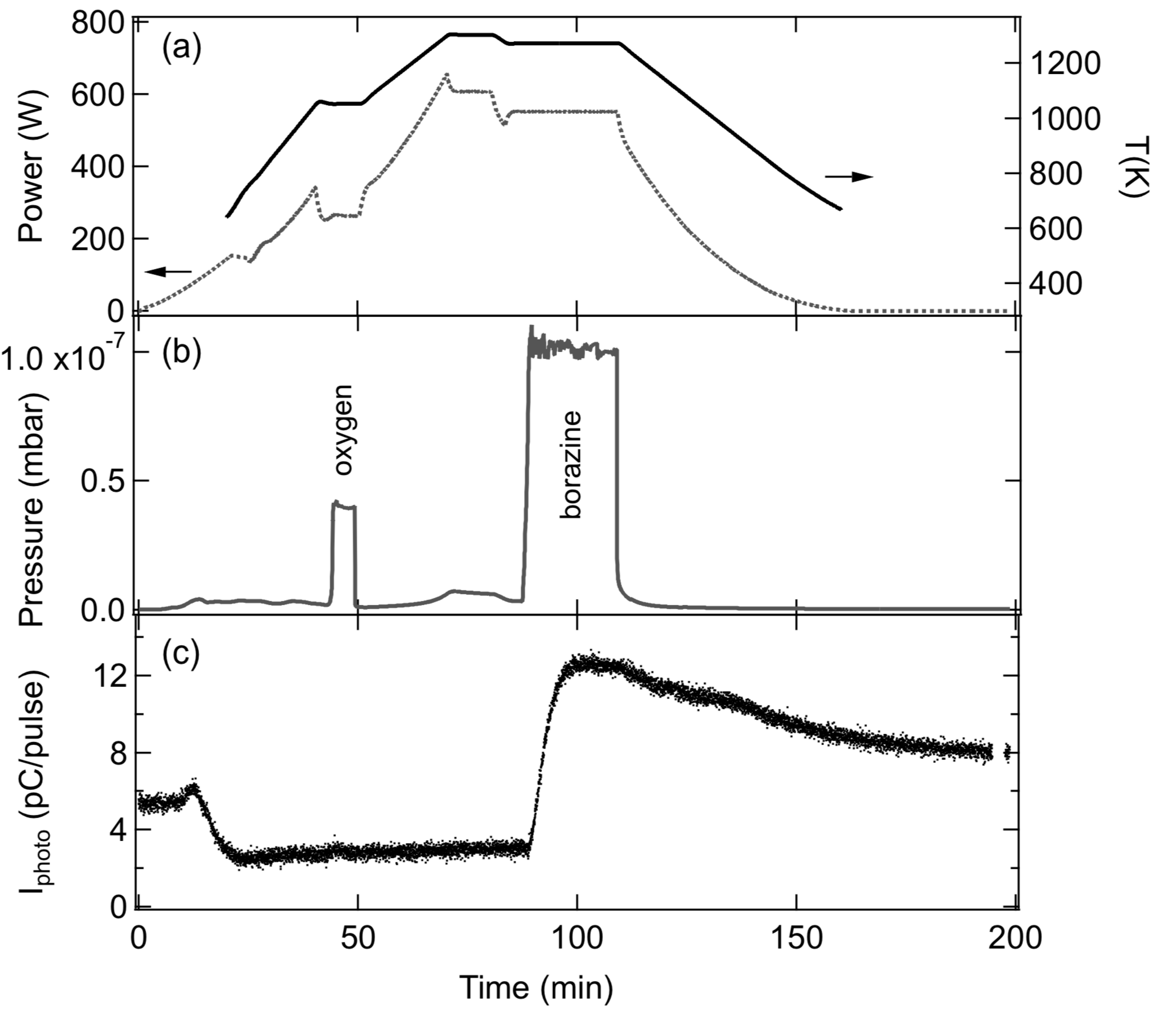}
  \caption{{\bf{Preparation protocol of a single layer $h$-BN on a Pt(111)/sapphire wafer.}} (a) Heater power (left axis) and sample temperature as read from the pyrometer (right axis). (b) Chamber pressure where oxygen and borazine exposure are marked. (c) Photocurrent that monitors the work function of the surface during the process.}
  \label{F1}
\end{figure}

Figure~\ref{F1} shows a typical 3 hours process-cycle for the production of a monolayer of hexagonal boron nitride on Pt(111).  The growth protocol includes temperature, pressure and surface status from photoelectron yield \cite{Hemmi20142}. The decrease of the yield upon heating to 700~K indicates desorption of gases from the surface.  The final oxygen exposure does not further decrease the yield at 1050~K as it is observed in the first oxygen annealing cycles and marks a clean Pt sample. At this temperature and oxygen pressure the equilibrium oxygen coverage is expected to be low and the work function should not strongly increase \cite{Derry1985}. The very small increase of the yield  upon the O$_2$ exposure at 1050~K may be due to an increased density of single Pt atoms on the surface or due to an increase of the background signal. The significant increase of the photocurrent followed by the admittance of  10$^{-7}$~mbar (HBNH)$_3$ during 20 minutes at 1273~K illustrates the formation of a $h$-BN layer. The decrease of the yield after preparation is also due to the decrease in sample temperature that has, together with the work function an influence on the photoelectron yield \cite{DuBridge1933}. In the following we will analyze the photoelectron yield curve in more detail and show on how much information on the nucleation and growth process may be retrieved.

The electron yield contains contributions from the $h$-BN and the platinum surface portion and a background from photoelectrons of the scattered light $Y_{{\rm{bkg}}}$. 
Assuming $Y_{{\rm{bkg}}}$ from the cold walls of the reactor to be constant, knowing the yield of the clean Pt $Y_0$ ($\Theta$=0) and the full monolayer $Y_1$ ($\Theta$=1) it is straight forward to determine the coverage $\Theta(Y)$:
\begin{equation}
\Theta = \frac{Y-Y_0}{Y_1-Y_0}
\label{Elamp1}
\end{equation}
In brackets we note that if $\Theta$ shall be determined for different temperatures $Y_0$ and $Y_1$ for the corresponding temperatures have to be known.

\begin{figure}
\includegraphics[width = 0.6\textwidth]{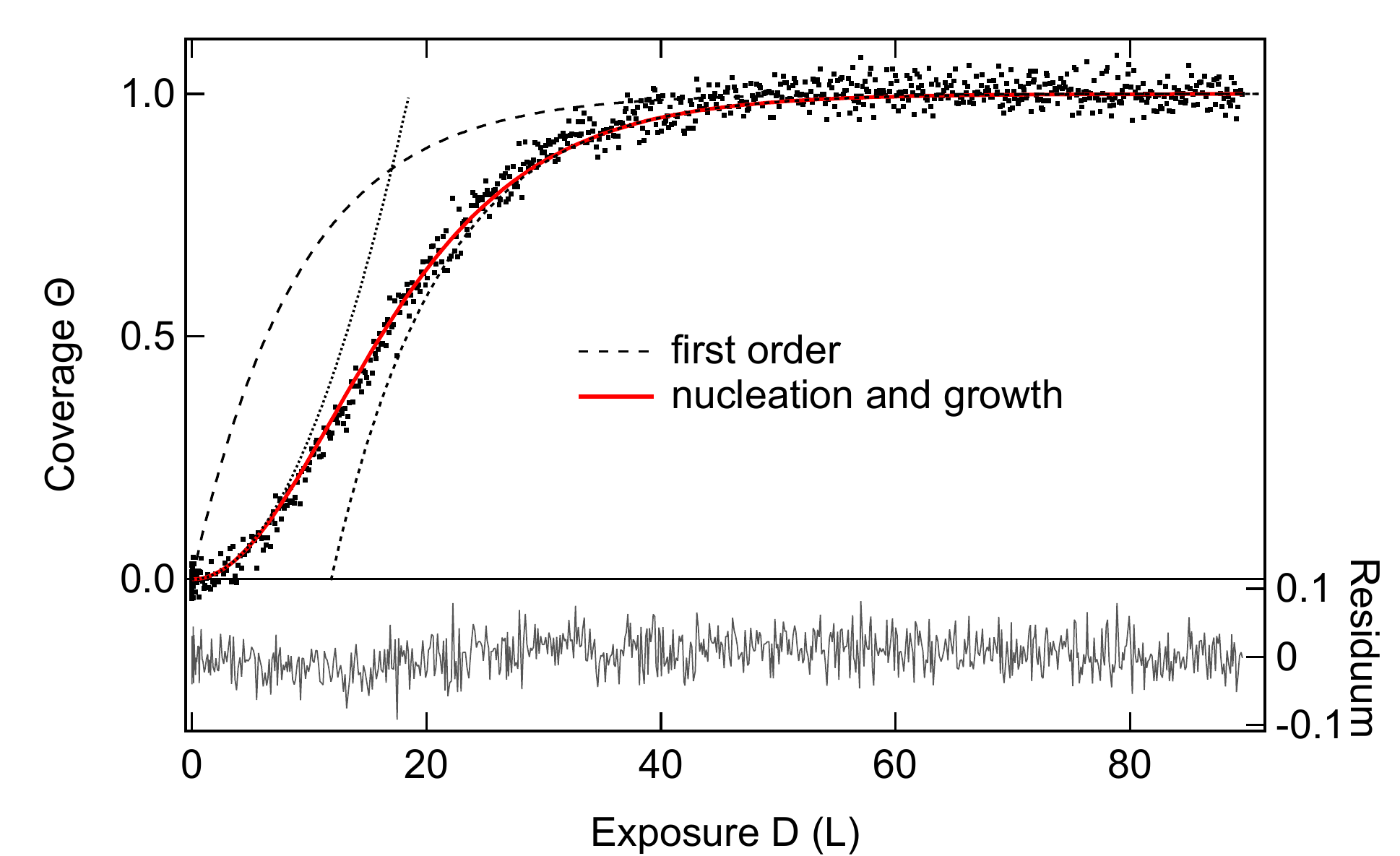}
  \caption{{\bf{$h$-BN growth kinetics on Pt(111).}} Top panel: Coverage $\Theta$ as inferred from the photoelectron yield vs. exposure $D$ in units of Langmuir (1L= 10$^{-6}~{\rm{Torr}}\cdot s)$. The borazine pressure was set to 10$^{-7}$ mbar and the sample temperature was 1273 K. The black dashed line mimics the first order model of Eqn. \ref{E2}, the red solid line the $\tanh^2$ nucleation and growth model of Eqn. \ref{E4}. The dotted lines indicate the asymptotic behaviour of the $\tanh^2$ function, like the initial growth $\propto D^2$ and the exponential approach of saturation coverage. Bottom panel residuum between experiment and nucleation and growth model. }
  \label{F2}
\end{figure}

Figure~\ref{F2} shows the {\mbox{$h$-BN}} coverage $\Theta$ on Pt(111) as derived from Eqn. \ref{Elamp1} versus borazine exposure $D(t)=\int_0^t p_{\rm{borazine}}(t') dt'$. 
In a picture where the growth rate is proportional to the non-covered surface $(1-\Theta)$, the sticking probability $s$ and the precursor gas pressure $p$, a first order rate model for the description of the growth of the monolayer ($\Theta=1$) applies:
\begin{equation}
d\Theta=(1-\Theta)  s~\frac{j_c}{n}~dt~,
\label{E1}
\end{equation}
where  $n$ is the areal density of BN units in $h$-BN (18.4 nm$^{-2}$) and
$j_c$ the collision current density of the incoming precursor molecules that is proportional to their pressure $p$ in the gas phase. For a borazine pressure of 10$^{-7}$ mbar and room temperature $j_c$  gets 0.17 borazine molecules or {\mbox{0.51 BN units nm$^{-2} s^{-1}$}}.

Substitution of $p\,dt$ with $dD$  and integration of Eq.\ref{E1} yields with the initial condition $\Theta (0)=0$:
\begin{equation}
\Theta(D)=1-\exp\left(-s~\frac{D}{D_0}\right)  ,
\label{E2}
\end{equation}
with the  constant $D_0$ being the precursor dose for getting a full monolayer if all incoming molecules would react to $h$-BN: $D_0= \frac{n}{3} \sqrt{2\pi \, m \, k_BT}$ = 2.7~Langmuir, with the mass $m$ and the temperature $T$ of the precursor gas. The factor $\frac{1}{3}$ applies for borazine that carries 3 BN units.

For larger coverages $\Theta > 0.5$ the growth rate slows down as expected from Eq.~\ref{E2} \cite{Sutter2011}, though it misses the correct scaling of the growth onset. 
Instead of $\Theta \propto D$ we observe a parabolic onset $\Theta \propto D^2$, which is indicative of nucleation and growth. 
An ansatz as 
\begin{equation}
d\Theta=(1-\Theta) \sqrt{\Theta/\Theta_0}~s~\frac{j_c}{n}~dt
\label{E3}
\end{equation}
provides a parabolic onset of growth.
The $\sqrt{\Theta}$ term models a growth proportional to the perimeter of the $h$-BN nuclei and complies with a picture where adsorbed precursor material may desorb after a certain residence time on the surface. 
The parameter $\Theta_0$ warrants dimensional homogeneity and accounts e.g. for the nucleation seed geometry and density.
Mass conservation asks $\Theta_0$ to be $\geq 1$, since the growth rate may not exceed $s (1-\Theta) j_c$. 
The validity of Eqn.~\ref{E3} may be challenged for large coverages, where e.g. percolation of the islands occurs and where the growth rate should not depend on $\sqrt{\Theta}$ - but over all, the observed kinetics fit well to this equation.

\begin{figure}
\includegraphics[width = 0.5\textwidth]{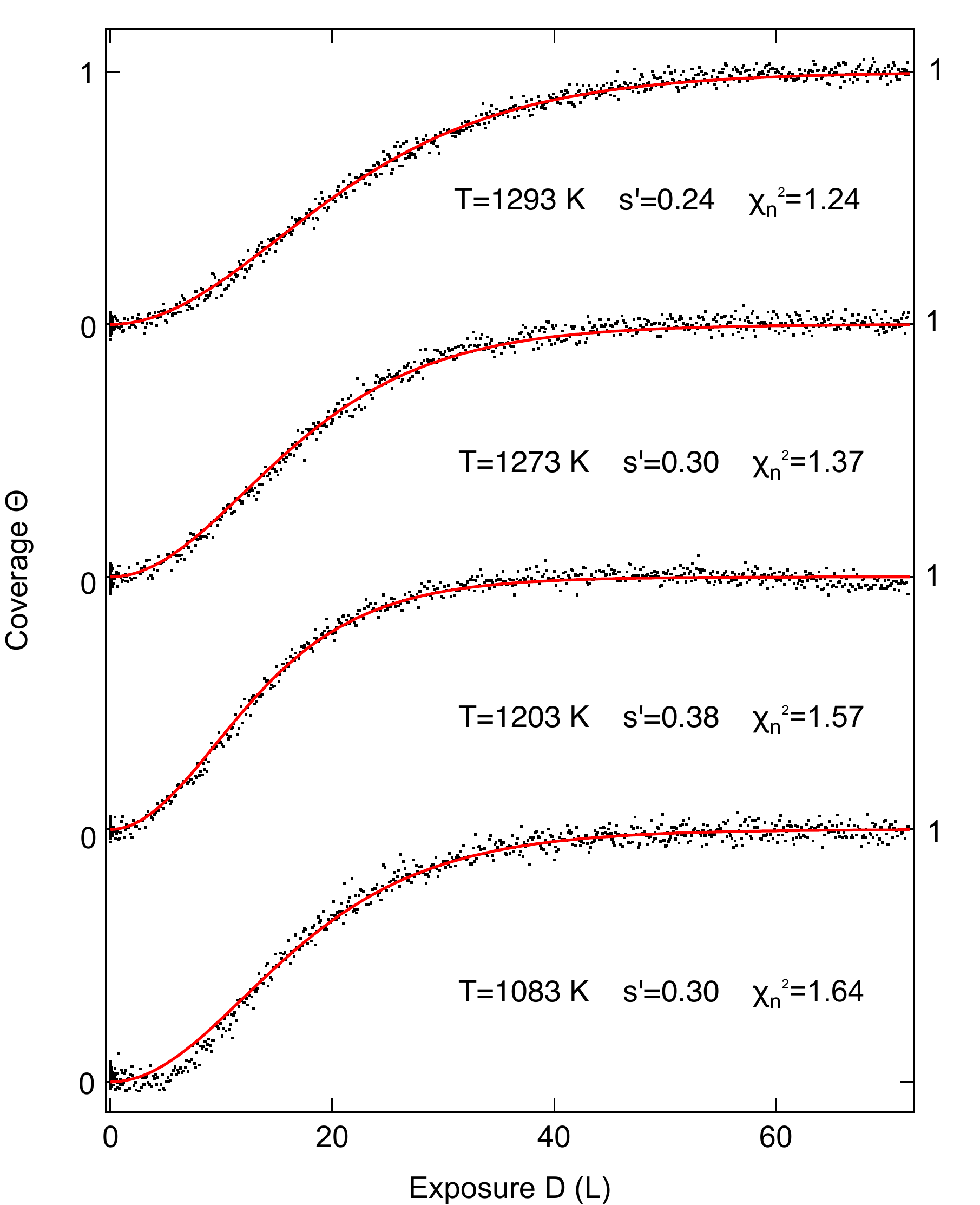}
  \caption{{\bf{Growth of $h$-BN on Pt(111) at different temperatures.}} Coverage vs. exposure to 10$^{-7}$ mbar borazine as inferred from photoelectron yield measurements for four different temperatures. Black dots data, red lines $\tanh^2$ fits with corresponding effective sticking parameters $s'$. $\chi_n^2$ is the $\chi^2$ of the fits normalized with the number of data points and the standard deviation of single yield measurements before borazine admission. A value $\chi_n^2$ of 1 indicates a perfect fit.}
  \label{F22}
\end{figure}

Substitution of $p\,dt$ with $dD$ and integration of Eqn.~\ref{E3} yields with the initial condition $\Theta(0)=0$ the hyperbolic tangent square expression:
\begin{equation}
\Theta(D)=\tanh^2{\left(\frac{s}{2\sqrt{\Theta_0}}\frac{D}{D_0} \right)}  .
\label{E4}
\end{equation}
It is seen that for small exposures the coverage $\Theta$ increases $\propto D^2$, and approaches full coverage with an exponential like in Eqn. \ref{E2}, but with an offset.
Figure \ref{F2} shows the almost perfect fit and the two asymptotes to one data set. The model provides a single parameter $s/\sqrt{\Theta_0}$=0.3 that characterizes the growth process. 
At this stage $s$ and $\Theta_0$ may not be separated and it is convenient to define the parameter $s'=s/\sqrt{\Theta_0}$ as the effective sticking coefficient.
For coverages approaching one we expect all precursor material that sticks on the surface to contribute to the growth, since it reaches the rim of a $h$-BN patch before desorption and this would imply $\Theta_0=1$. Analyzing the data sets from different growth temperatures in Figure \ref{F22} with an extended model where $\Theta_0$ is not constant but approaches one exponentially, shows $\Theta_0$ to be larger than one for all exposures and that the $\chi^2$ values of the fits further decrease by about 10~\%. Corresponding initial $\Theta_0(D=0)$ values carry additional information, but here we refrain from refining the model of Eqn. \ref{E4}.  

In Figure \ref{F22} four subsequent preparations, each starting from clean Pt(111) but at different growth temperatures are shown. 
It can be seen that $s'$ is highly non-linear with temperature, and does not decrease monotonously with preparation temperature.
This indicates different growth regimes and layer qualities.
We furthermore note that the goodness of fit increases with increasing temperature. 
As a measure for the goodness we use $\chi_n^2$ which is the $\chi^2$ normalized with the number of data points and the standard deviation of single yield measurements before borazine admission. Large $\chi_n^2$ indicate a bad fit and a perfect fit is reached with $\chi_n^2$ =1.

Comparison of different photoemission yield measurements provides a direct measure for the reproducibility of a process and
allows to forecast the expected quality of the grown material. The course of the work function that is derived from the yield is a quality measure that still may be equivocal. For validation it should be correlated with other means of quality control. In the next section we apply the inspection of the crystal-quality after growth by scanning low energy electron diffraction (x-y LEED).

\subsection{x-y LEED}

\begin{figure}
\includegraphics[width = 0.9\textwidth]{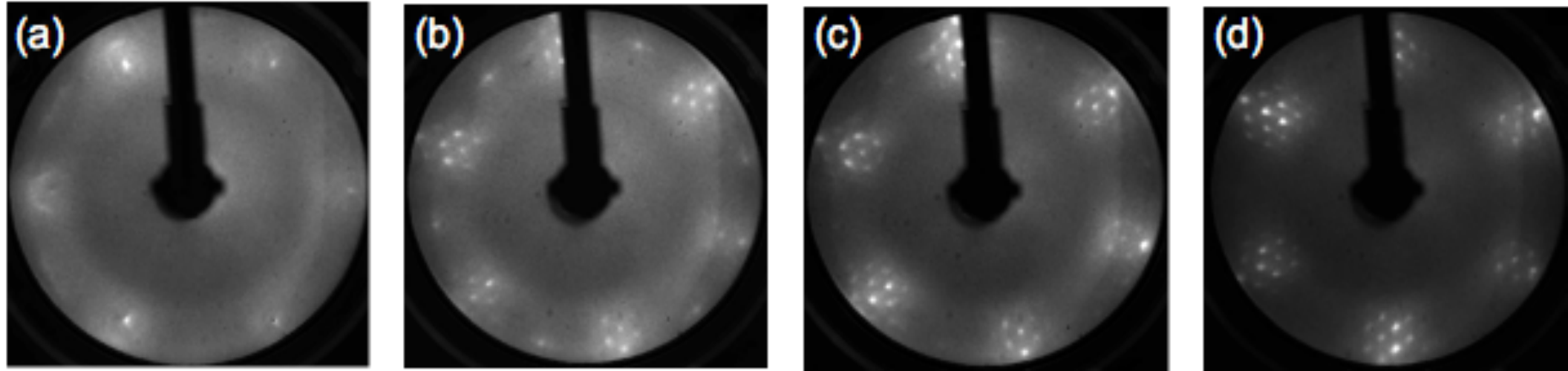}
  \caption{{\bf{LEED for subsequent $h$-BN preparations on Pt(111).}} Electron energy E = 100 eV. (a) T = 1083 K, (b) T = 1203 K. Note the minority $h$-BN phase that is rotated 30 degrees away from the high symmetry substrate direction, (c) T = 1273 K, (d) T = 1293 K.}
  \label{F3}
\end{figure}

After each preparation where the kinetics are shown in Figure \ref{F22} the wafer surface was investigated by LEED before it was cleaned again for the next growth cycle. Figure~\ref{F3} shows LEED data after the preparations of single layer $h$-BN on Pt(111). The dominant LEED pattern corresponds to an aligned 10 on 9 superstructure \cite{Paffett1990}. At the growth temperatures of T= 1083 K and 1203 K a minority $h$-BN phase that is rotated 30 degrees away from the high symmetry substrate direction is found. This reminds to the earlier study where trichloroborazine was used as a precursor  \cite{Muller2005}. Furthermore, at low temperature a "moir\'e-ring" with the $h$-BN lattice constant indicates azimuthally disordered $h$-BN. Such moir\'e-rings were also observed for $h$-BN/Pd(111) \cite{Morscher2006}. However, with increasing preparation temperature the aligned 10 on 9 structure dominates as shown by Preobrajenski {\it{et al.}} \cite{Preo2007, Cavar2008}. In the case of the highest growth temperature of T = 1293 K, rotational misalignment disappeared below the detection limit of our LEED instrument. This indicates that aligned $h$-BN may be grown on Pt(111) with a very high crystal-quality, if the growth is performed at high temperatures.

\begin{figure}
\includegraphics[width = 0.9\textwidth]{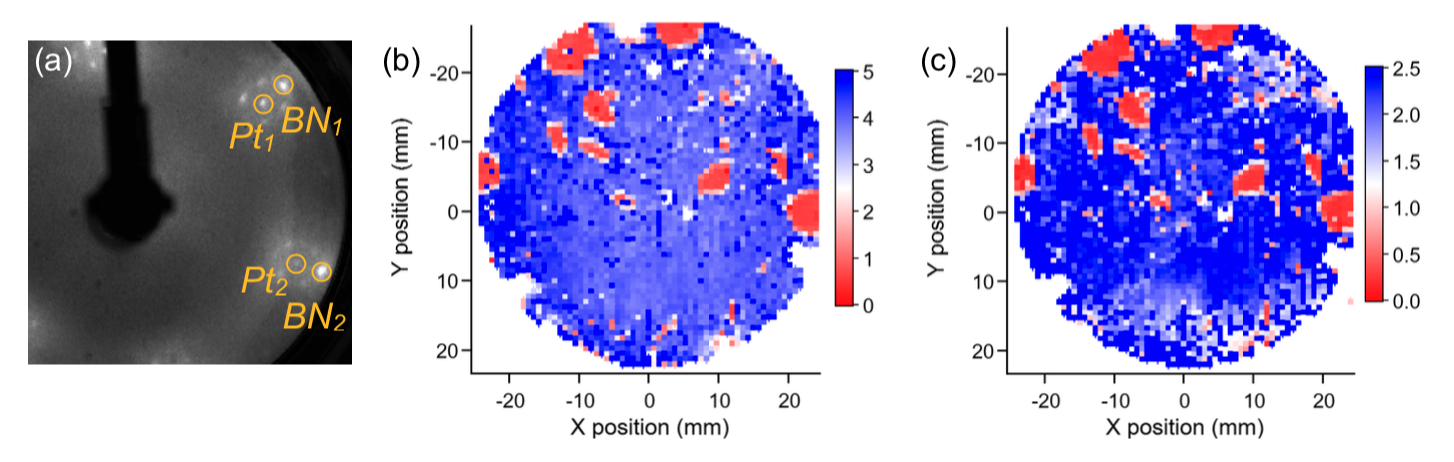}
  \caption{{\bf{x-y LEED scans for $h$-BN on Pt(111).}} Preparation at 1273~K. E=100~eV. (a) Zoom into the LEED pattern of Figure \ref{F3}(c). The $Pt_1$, $Pt_2$, $BN_1$ and $BN_2$ spots for twin analysis are indicated. (b-c) x-y LEED scans with 4400 data points across the wafer reveal twin locations. (b) Intensity ratio between Pt principle LEED spots $I(Pt_1):I(Pt_2)$. (c) Intensity ratio between $h$-BN principle spots $I(BN_1):I(BN_2)$.}
  \label{F4}
\end{figure}

In order to quantify the crystal-quality of the $h$-BN on the wafer scale, the sample was scanned in front of the LEED optics on an x-y piezo motor driven stage. 
In Figure \ref{F4} the results of 4400 sequential LEED measurements for the preparation at 1273 K are shown. The whole wafer displays a 10 on 9 superstructure LEED pattern. From the orientation of the LEED patterns we infer millimeter sized twin-domains. At the electron energy of 100~eV the diffraction patterns from $fcc$ Pt(111) and $h$-BN are three fold symmetric and the pattern of the corresponding twin domain is rotated by 180 degree. Therefore, the intensity ratios of diffraction spots, as indicated in Figure \ref{F4}(a) are expected to be different for different domains. The $Pt_1:Pt_2$ intensity ratios in Figure \ref{F4}(b) indicate a twinning that stems from the substrate growth process. The $BN_1:BN_2$ ratios in Figure \ref{F4}(c) follow those of the substrate. At the length scale of the experiment (500 $\mu$m) we find no twinned $h$-BN on a given Pt(111) orientation. For the case of $h$-BN/Ni(111) such $fcc$ and $hcp$ domains have been found \cite{Auwarter2003}, while recent $\mu$-LEED experiments showed as well the formation of  $fcc$ and $hcp$ $h$-BN domains on Rh(111) on the micrometer scale \cite{Cun2020}.

\section{Conclusions}
We report the observation of the growth kinetics of single layer $h$-BN on 2 inch Pt(111)/sapphire wafers with in-situ photoelectron yield measurements.
The coverage vs. exposure curves may be well described with a $\tanh^2$ nucleation and growth model that indicates different effective sticking coefficients for different temperatures.
Such growth monitoring data allow an immediate control and improvement of the growth process.
After growth the $h$-BN/Pt(111) layers are checked with scanning x-y LEED that reveals the  $h$-BN quality and twin-domains in the substrate. The best crystal-quality $h$-BN on Pt(111) is obtained at growth temperatures above 1000 $^\circ$C.
\begin{acknowledgement}
Financial support by the European Commission under the Graphene Flagship (contract no.  CNECT-ICT-604391) is  gratefully acknowledged. 
AH acknowledges a Forschungskredit of the University of Z\"urich (grant no. FK-20-206 114).

\end{acknowledgement}

\providecommand{\latin}[1]{#1}
\makeatletter
\providecommand{\doi}
  {\begingroup\let\do\@makeother\dospecials
  \catcode`\{=1 \catcode`\}=2 \doi@aux}
\providecommand{\doi@aux}[1]{\endgroup\texttt{#1}}
\makeatother
\providecommand*\mcitethebibliography{\thebibliography}
\csname @ifundefined\endcsname{endmcitethebibliography}
  {\let\endmcitethebibliography\endthebibliography}{}

\end{document}